\documentclass[prl,aps,twocolumn,preprintnumbers,superscriptaddress]{revtex4}

\usepackage{mathrsfs}
\usepackage{amssymb}
\usepackage{graphicx}
\usepackage{dcolumn}
\usepackage{bm}
\usepackage{graphicx}
\usepackage{float}
\usepackage{longtable}
\usepackage{amsmath}
\usepackage{color}
\usepackage{multirow}
\usepackage[ulem=normalem]{changes}

\newfont{\largemi}{cmmi10}
\newfont{\smallmi}{cmmi6}

\draft

\def\eqref#1{Eq.~(\ref{eq:#1})}

\begin{document}

\title{Nucleon-pair coupling scheme in Elliott's SU(3) model}

\author{G. J. Fu\footnote{gjfu@tongji.edu.cn}}
\affiliation{School of Physics Science and Engineering, Tongji University, Shanghai 200092, China}

\author{Calvin W. Johnson\footnote{cjohnson@sdsu.edu}}
\affiliation{Department of Physics, San Diego State University, 5500 Campanile Drive, San Diego, CA 92182-1233}

\author{P. Van Isacker\footnote{isacker@ganil.fr}}
\affiliation{Grand Acc\'{e}l\'{e}rateur National d'Ions Lourds, CEA/DRF-CNRS/IN2P3, Boulevard Henri Becquerel, F-14076 Caen, France}

\author{Zhongzhou Ren\footnote{zren@tongji.edu.cn}}
\affiliation{School of Physics Science and Engineering, Tongji University, Shanghai 200092, China}

\date{\today}

\begin{abstract}
Elliott's SU(3) model is at the basis of the shell-model description
of rotational motion in atomic nuclei.
We demonstrate that SU(3) symmetry can be realized
in a truncated shell-model space
if constructed in terms of a sufficient number of collective $S$, $D$, $G$, \dots pairs
(i.e., with angular momentum zero, two, four, \dots)
and if the structure of the pairs is optimally determined
either by a conjugate-gradient minimization method
or from a Hartree-Fock intrinsic state.
We illustrate the procedure for 6 protons and 6 neutrons in the $pf$ ($sdg$) shell
and exactly reproduce the level energies and electric quadrupole properties
of the ground-state rotational band with $SDG$ ($SDGI$) pairs. 
The $SD$-pair approximation without significant renormalization, on the other hand, cannot describe the full SU(3) collectivity.
A mapping from Elliott's fermionic SU(3) model to systems with $s$, $d$, $g$, \dots bosons provides insight into the existence of a decoupled collective subspace
in terms of $S$, $D$, $G$, \dots pairs.

\end{abstract}



\maketitle

Atomic nuclei exhibit a wide variety of behaviors,
ranging from  single-particle motion
to superconducting-like pairing
to vibrational and rotational modes.
To a large extent the story of nuclear structure
is the quest to encompass the widest range of behaviors
within the fewest degrees of freedom.
In the early stage of nuclear physics,
the spherical nuclear shell model~\cite{Mayer,Jensen}
stressed the single-particle nature of the nucleons in a nucleus,
while the geometric collective model~\cite{BM1,BM2}
and the Nilsson mean-field model~\cite{Nilsson}
pointed the way to describing rotational bands
by emphasizing permanent quadrupole deformations~\cite{Rainwater} in ``intrinsic'' states.
The reconciliation between these \mbox{two} pictures
has been one of the most important advances in our understanding of the structure of nuclei.
It was in large part due to Elliott who showed,
on the basis of an underlying SU(3) symmetry,
how to obtain deformed ``intrinsic'' states
in a finite harmonic-oscillator single-particle basis
occupied by nucleons that interact through a quadrupole-quadrupole force~\cite{Elliott58}.
This major step forward provided a microscopic interpretation
of rotational motion in the context of the spherical shell model
and, more recently, led to the symmetry-adapted no-core shell model~\cite{symmetryadapted}.

Although the spherical shell model does provide a general framework
to reproduce rotational bands~\cite{Caurier05} and shape coexistence~\cite{Heyde11} in light- and medium-mass nuclei,
it is computationally still extremely challenging to describe deformation in heavier-mass regions~\cite{Otsuka19}.
Approximations must be sought.
A tremendous simplification of the shell model occurs
by considering only pairs of nucleons with angular momentum 0 and 2,
and treating them as ($s$ and $d$) bosons.
This approximation, known as the interacting boson model (IBM)~\cite{IBM1,IBM2},
is particularly attractive because of its symmetry treatment in terms of a U(6) Lie algebra,
which allows a spherical U(5), a deformed SU(3), and an intermediate SO(6) limit.
While the IBM has been connected to the shell model for spherical nuclei~\cite{OAI,GJ95},
such relation has never been established for deformed nuclei,
in which case the IBM has rather been derived
from mean-field models~\cite{Nomura1,Nomura2}.

The nucleon-pair approximation (NPA)~\cite{NPA1,NPA2}
is one possible truncation scheme of the shell-model configuration space. 
The building blocks of the NPA are fermion pairs with certain angular momenta.
Calculations are carried out in a fully fermionic framework,
albeit in a severely reduced model space
defined by the most important degrees of freedom in terms of pairs.
The NPA therefore can be considered as intermediate
between the full-configuration shell model
and models that adopt the same degrees of freedom
as the nucleon pairs but in terms of bosons. 
While the NPA has been successful for nearly spherical nuclei~\cite{NPAr,gs1,gs2,gs3,bpa1,bpa2,gs4,Lei}, 
previous studies for well-deformed nuclei are not satisfactory.
For example, in the fermion dynamical symmetry model~\cite{FDSM1,FDSM2}
an SU(3) limit with Sp(6) symmetry can be constructed in terms of $S$ and $D$ pairs
but their symmetry-determined structure
is far removed from that of realistic pairs~\cite{Halse89}.
Also, the binding energy, moment of inertia, and electric quadrupole ($E2$) transitions
calculated in an $SD$-pair approximation
are much smaller than those obtained
in Elliott's SU(3) limit for the $pf$ and $sdg$ shells~\cite{Zhao2000}.

In this Letter we successfully apply the NPA of the shell model to well-deformed nuclei.
We show that the low-energy excitations of many-nucleon systems in Elliott's SU(3) limit
can be exactly reproduced with a suitable choice of pairs in the NPA.
We obtain an understanding of this observation
through a mapping to a corresponding boson model.

We consider an example system with even numbers of protons and neutrons
in a degenerate $pf$ or $sdg$ shell,
interacting through a quadrupole-quadrupole force of the form,
\begin{eqnarray}\label{QQ}
V_Q = -(Q_{\pi} + Q_{\nu}) \cdot (Q_{\pi} + Q_{\nu}),
\end{eqnarray}
where $Q_{\pi}$ ($Q_{\nu}$) is the quadrupole operator for protons (neutrons),
\begin{eqnarray}
Q =  - \displaystyle\sum_{\alpha \beta} \displaystyle
\frac{ \langle n_{\alpha}l_{\alpha}j_{\alpha}  \| r^2 Y_2 \| n_{\beta} l_{\beta}j_{\beta} \rangle }{\sqrt{5 }r_0^2}
 \left( a_{\alpha}^{\dagger} \times \tilde{a}_{\beta} \right)^{(2)} .
\end{eqnarray}
Greek letters $\alpha$, $\beta,\ldots$ denote harmonic-oscillator single-particle orbits labeled by $n$, $l$, $j$, and $j_z$; 
$a_{\alpha}^{\dagger}$ and $\tilde{a}_{\beta}$ are the nucleon creation operator and its time-reversed form for the annihilation operator, respectively;
and $r_0$ is the harmonic-oscillator length.
As shown in Ref.~\cite{Elliott58}, the interaction $V_Q$
is a combination of the Casimir operators of SU(3) and SO(3),
and its eigenstates are therefore classified
by (irreducible) representations of these algebras
with eigenenergies given by
\begin{eqnarray}\label{solvable}
-\frac{5}{2\pi} \left[ \frac{1}{2}(\lambda^2 + \lambda\mu + \mu^2 + 3\lambda + 3\mu) - \frac{3}{8}L(L+1) \right],
\end{eqnarray}
in terms of the SU(3) labels $(\lambda,\mu)$
and the SO(3) label $L$, the total orbital angular momentum.
Several useful SU(3) representations for low-lying states
can be found in Ref.~\cite{Zhao2000}.

In the following we discuss in detail the case of 6 protons and 6 neutrons (6p-6n)
in the NPA of the shell model
and subsequently generalize to other numbers of nucleons.
A nucleon-pair state of 6 protons is written as
\begin{eqnarray} \label{protonbasis}
| \varphi^{(I_{\pi})} \rangle &=& \left( ( {{A}^{({J}_{1})}}^{\dag}\times {{A}^{({J}_{2})}}^{\dag}  )^{(I_2)} \times {{A}^{({J}_{3})}}^{\dag} \right)^{(I_{\pi})}  |0\rangle ,
\end{eqnarray}
where $I_2$ is an intermediate angular momentum and ${{A}^{(J)}}^{\dag}$ is the creation operator of a collective pair with angular momentum $J$:
\begin{eqnarray} \label{pair}
{{A}^{(J)}}^{\dag}
= \sum_{{\alpha} \leq {\beta} } y_{J}({\alpha} {\beta})  \left( a_{{\alpha}}^{\dagger} \times a_{{\beta}}^{\dagger} \right)^{(J)},
\end{eqnarray}
where $y_{J}({\alpha} {\beta})$ is the pair-structure coefficient.
For systems with protons and neutrons, we construct the basis by coupling the proton and neutron pair states to a state with total angular momentum $I$, i.e.,
$ | \psi^{(I)} \rangle  = \left( | \varphi^{(I_{\pi})} \rangle \times | \varphi^{(I_{\nu})} \rangle \right)^{(I)} $.
Level energies and wave functions are obtained
by diagonalization of the Hamiltonian matrix
in the space spanned by $\left\{ | \psi^{(I)} \rangle  \right\}$,
that is, from a configuration-interaction calculation.
If a sufficient number of pair states are considered in Eq.~(\ref{protonbasis}),
the NPA model space can be made exactly
equivalent to the full shell-model space.
The interest of the NPA, however, is to restrict to the relevant pairs
and describe low-energy nuclear structure in a truncated shell-model space.

The selection of relevant pairs with the correct structure in Eq.~(\ref{pair})
has been a long standing problem in NPA calculations.
Recent applications choose pairs by the generalized seniority scheme (GS).
Specifically, one optimizes the structure coefficients of the $S$ pair
by minimizing the expectation value of the Hamiltonian in the $S$-pair condensate
and one obtains other pairs by diagonalizing the Hamiltonian matrix
in the space spanned by GS-two (i.e., one-broken-pair) states~\cite{gs2,Xu2009}.
The collective pairs obtained with the GS approach
provide a good description of nearly-spherical nuclei but,
as recognized in Ref.~\cite{Lei12un} and as will also be shown below,
they are inappropriate in deformed nuclei.
Instead we use the conjugate gradient (CG) method~\cite{CG1,CG2},
where the structure coefficients of {\em all} pairs considered in the basis
are simultaneously optimized by minimizing the ground-state energy
in a series of iterative NPA calculations for a given Hamiltonian.
The initial pairs in this iterative procedure are SU(3) tensors,
obtained by diagonalizing $V_Q$ in a two-particle basis and retaining the lowest-energy pair.

\begin{figure}
	\includegraphics[width = 0.48 \textwidth]{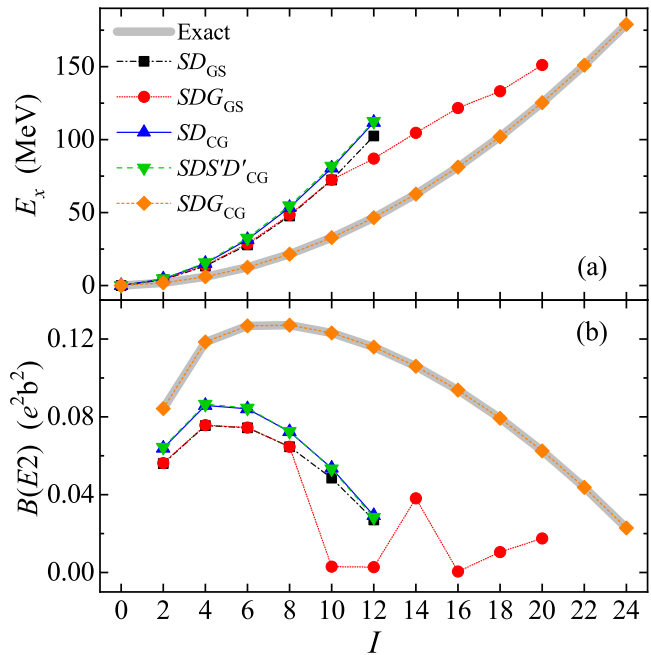}
	\caption{\label{fig1}
	(a) Excitation energy and
	(b) electric quadrupole reduced transition probability $B(E2;I \rightarrow I-2  )$
	for the ground rotational band of 6 protons and 6 neutrons
	in the $pf$ shell in Elliott's SU(3) model.
	The subscript ``GS'' stands for generalized seniority and ``CG'' for conjugate gradient (see text). 
	}
\end{figure}

\begin{figure}
	\includegraphics[width = 0.48 \textwidth]{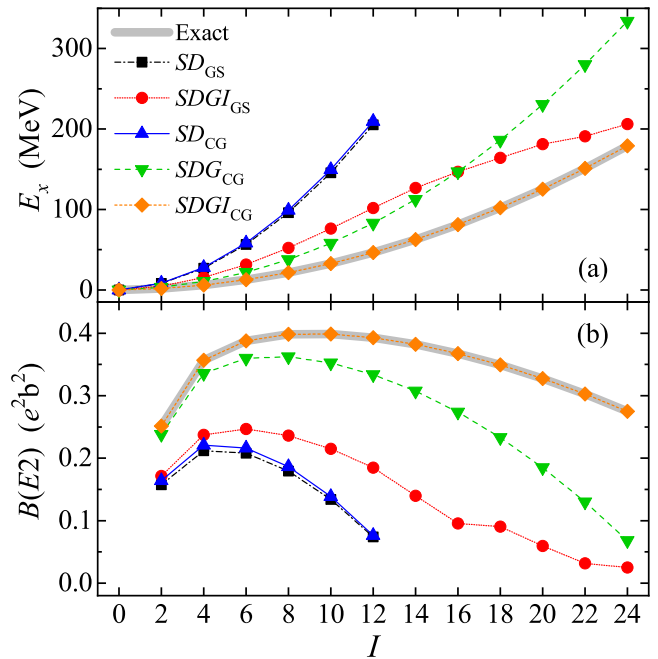}
	\caption{\label{fig2} Same as Fig. \ref{fig1} for the $sdg$ shell.
	}
\end{figure}

Figure \ref{fig1} shows, for a 6p-6n system in the $pf$ shell,
the results of various NPA calculations
concerning excitation energies
and $E2$ reduced transition probabilities
(with the standard effective charges $e_{\pi}=1.5$ and $e_{\nu}=0.5$)
for the lowest rotational band.
These are compared to the exact results of Elliott's model,
where the ground band belongs to the SU(3) representation $ (\lambda,\mu) = (24,0)$.
Surprisingly, the $SDG$-pair approximation of the shell model
in the CG approach (denoted as $SDG_{\rm CG}$)
reproduces the exact binding energy,
$810/\pi$ MeV according to Eq. (\ref{solvable}),
to a precision of eight digits,
as well as the exact excitation energies for the entire ground band. 
One can understand the occurrence of the $ (24,0)$ representation
from the coupling of $(12,0)$ for the six protons and six neutrons separately
and, in fact, all bands contained in the product $(12,0)\times(12,0)$,
i.e.\ $(24,0)$, $(22,1)$,\dots,$(0,12)$,
are exactly reproduced in the $SDG_{\rm CG}$-pair truncated space.
We also find that the results of the $SDG$-pair approximation
are close to the exact results if the pairs are SU(3) tensors.
For example, with such pairs the calculation reproduces 98\% of the exact binding energy,
99\% of the exact moment of inertia, and 97\% of the exact $B(E2)$ values.

On the other hand, the results of the $SDG$-pair approximation deteriorate
if the pairs are obtained with the GS approach (denoted as $SDG_{\rm GS}$),
which reproduces only 76\% of the exact binding energy.
Furthermore, $SDG_{\rm GS}$ fails to describe the quadrupole collectivity:
The moment of inertia predicted by $SDG_{\rm GS}$
is only $\sim$43\% of the exact one,
the predicted $B(E2)$ values are too small,
and the yrast states with angular momentum $I\geq10$
do not follow the behavior of a quantum rotor.
One concludes that the structure of the collective pairs,
as determined by the GS approach,
is not suitable for the description of well-deformed nuclei.

It is also of interest to investigate the standard $SD$-pair approximation of the shell model
and results of the $SD_{\rm GS}$-, $SD_{\rm CG}$-, and $SDS^{\prime}D^{\prime}_{\rm CG}$-pair approximations
are shown in Fig. \ref{fig1}.
Here $S^{\prime}$ and $D^{\prime}$ are collective pairs with angular momentum 0 and 2
but orthogonal to the $S$ and $D$ pairs, respectively.
While the CG approach provides the numerically optimal solution
in $SD_{\rm CG}$- and $SDS^{\prime}D^{\prime}_{\rm CG}$-pair approximations,
the results nonetheless are underwhelming.
In the $SD_{\rm GS}$, $SD_{\rm CG}$, and $SDS^{\prime}D^{\prime}_{\rm CG}$ spaces
only 76\%, 83\%, and 84\% of the exact binding energy are reproduced, respectively, 
and the predicted moments of inertia and $B(E2)$ strengths
are evidently smaller than the exact SU(3) results.
We conclude that the collective $SD$ pairs
cannot fully explain the quadrupole collectivity of the SU(3) states.
Interestingly, the excitation energies of the yrast states
predicted by the $SD$-pair approximations follow an $I(I+1)$ rule
and the $B(E2)$ strength exhibits a nearly-parabolic shape [see Fig. \ref{fig1}(b)],
two typical features of rotational motion.
This raises the hope that an {\em effective} Hamiltonian
and {\em effective} charges can be derived in the restricted $SD_{\rm CG}$ space,
which takes into account the coupling with the excluded space.
This conclusion is in line with a more phenomenological approach~\cite{Nomura2},
in which an $L \cdot L$ term is added to the Hamiltonian,
such that properties of low-lying states of well-deformed nuclei
are reproduced in $sd$-IBM.

Figure \ref{fig2} shows the corresponding results of for the 6p-6n system in the $sdg$ shell.
In this case the $SDGI_{\rm CG}$-pair approximation of the shell model
reproduces exactly the SU(3) results
and all states belonging to the coupled representation $(18,0)\times(18,0)$,
i.e.\ $(36,0)$, $(34,1)$,\dots,$(0,18)$,
are fully contained in the $SDGI_{\rm CG}$-pair truncated space.
Again, if the pairs are SU(3) tensors, the $SDGI$-pair approximation is close to the exact result
and reproduces 99\% of the exact binding energy, 97\% of the exact moment of inertia, and 99\% of the exact $B(E2)$ values.
The $SDG_{\rm CG}$-pair approximation
yields 96\% of the binding energy and 57\% of the moment of inertia.
The predicted $B(E2)$ strength in the $SDG_{\rm CG}$-pair approximation
is close to the exact result for low angular momenta but deteriorates as angular momentum $I$ increases.
The necessity of renormalization is even larger 
in the $SD_{\rm CG}$-pair approximation.

Let us now try to understand the above results.
Specifically, why is it that the SU(3) results in the $pf$ shell
are exactly reproduced with $SDG$ but not with $SD$ pairs?
Similarly, why is it that SU(3) in the $sdg$ shell
cannot be represented with $SD$ or $SDG$
but requires $SDGI$ pairs?
To explain these findings, we
invoke a mapping to a system
with corresponding $s$, $d$, $g$, and $i$ bosons
(denoted as $sd$-, $sdg$-, or $sdgi$-IBM)
and the bosonic realization of SU(3).
The mapping is further specified by the fact
that the quadrupole-quadrupole interaction $V_Q$ is an SU(4) invariant
and, consequently, one aims to realize
the symmetries associated with Wigner's supermultiplet model~\cite{Wigner}
in terms of bosons.
An SU(4)-invariant boson model, known as IBM-4~\cite{Elliott81},
requires to assign to each boson
a spin-isospin of $(s,t)=(0,1)$ or $(1,0)$,
giving rise to a spin-isospin algebra ${\rm U}_{st}(6)$.

The SU(3) limit can be realized in terms of bosons
by first decoupling the orbital angular momentum 
from the spin-isospin of the bosons.
For an $n_b$-boson state this leads to the classification 
\begin{eqnarray} 
\begin{array}{ccccc}
{\rm U}(6\Lambda) & \supset & {\rm U}(\Lambda) & \otimes & {\rm U}_{st}(6) \\
\downarrow &     & \downarrow &   & \downarrow \\
 \left[ n_b \right] &   & \left[ \bar{h} \right] \equiv \left[ h_1,...,h_6 \right] & & \left[ \bar{h} \right] \equiv \left[ h_1,...,h_6 \right]
\end{array}  ,
\end{eqnarray}
with $\Lambda=6$, 15, and 28
for $sd$-, $sdg$-, and $sdgi$-IBM, respectively.
The six labels $[\bar h]$ are a partition of $n_b$
such that $h_1\geq h_2\geq\cdots\geq h_6$;
they specify the representations of ${\rm U}(\Lambda)$ and ${\rm U}_{st}(6)$,
which by virtue of the overall ${\rm U}(6\Lambda)$ symmetry of the bosons
must be identical.
For all above values of $\Lambda$ (i.e., $\Lambda=6$, 15, and 28),
Elliott's SU(3) appears as a subalgebra of ${\rm U}(\Lambda)$,
\begin{eqnarray} \label{subalg1}
\begin{array}{ccccccc} 
 {\rm U}( \Lambda)& \supset &{\rm U}(3)& \supset &{\rm SU}(3)& \supset &{\rm SO}(3)  \\
 \downarrow &   &\downarrow &   & \downarrow &  & \downarrow    \\
 \left[ \bar{h} \right]&     &\left[ h_1^{\prime\prime},h_2^{\prime\prime},h_3^{\prime\prime} \right]&    & (\lambda,\mu) &   K   &  L 
\end{array},
\end{eqnarray}
while Wigner's SU(4) occurs as a subalgebra of ${\rm U}_{st}(6)$,
\begin{eqnarray} \label{subalg2}
\begin{array}{ccccccc}
{\rm U}_{st}(6) & \supset & {\rm SU}_{st}(4) & \supset & {\rm SU}_{s}(2) &\otimes& {\rm SU}_{t}(2)  \\
\downarrow &   &\downarrow &   & \downarrow  &   & \downarrow      \\
\left[ \bar{h} \right]&   & (\lambda^{\prime},\mu^{\prime},\nu^{\prime}) & &  S  & &  T 
\end{array}.
\end{eqnarray}
The quantum numbers $(\lambda,\mu)$, $K$, and $L$ in Eq.~(\ref{subalg1})
and $(\lambda',\mu',\nu')$, $S$, and $T$ in Eq.~(\ref{subalg2})
have an interpretation identical to that
in Elliott's fermionic SU(3) model~\cite{Elliott58,Isacker2016}.

The SU(3) labels $(\lambda,\mu)$ in the different versions of the IBM
can be worked out with the following procedure~\cite{Elliott1999}.
For a given number of bosons $n_b$,
one enumerates all possible Young diagrams $[\bar h]$ of ${\rm U}( \Lambda)$ or $ {\rm U}_{st}(6)$. 
For each $[\bar h]$ one obtains the ${\rm SU}_{st}(4)$ labels $(\lambda',\mu',\nu')$
from the branching rule ${\rm U}(6)\supset{\rm SU}(4)$,
and retains only the ones that contain the favored supermultiplet.
Finally, the SU(3) labels $(\lambda,\mu)$ for the above $[\bar h]$ are found
from the ${\rm U}(\Lambda)\supset{\rm SU}(3)$ branching rule.

Let us apply this procedure to the 6p-6n system in the $pf$ shell.
The lowest eigenstates of the quadrupole-quadrupole interaction
belong to the favored SU(4) supermultiplet $(\lambda',\mu',\nu')=(0,0,0)$
and the leading (fermionic) SU(3) representation is $(\lambda,\mu)=(24,0)$.
For $n_b=6$ bosons, the ${\rm U}_{st}(6)$ or ${\rm U}(\Lambda)$ representations
containing this favored supermultiplet $(0,0,0)$
are $[\bar h]=[6]$, $[4,2]$, $[2^3]$, and $[1^6]$,
which have the SU(3) labels $(\lambda,\mu)$ as listed in Table~\ref{table1}
for the $sd$-, $sdg$-, and $sdgi$-IBM.
The leading SU(3) representation $(24,0)$ is not contained in $sd$-IBM
but is present in the $[6]$ representation of U(15),
and therefore it is contained in $sdg$-IBM.
Similarly, 6p-6n in the $sdg$ shell
give rise to the leading SU(3) representation $(36,0)$,
which is not contained in $sd$- nor $sdg$-IBM
but present in $sdgi$-IBM.

The generalization to the 2p-2n ($n=4$) and 4p-4n ($n=8$) systems
in the $pf$ and $sdg$ shells is summarized in Table~\ref{table2}.
The second column lists the leading fermionic SU(3) representations
and the third, fourth, and fifth columns indicate
whether this representation is contained in $sd$-, $sdg$-, and $sdgi$-IBM, respectively.
A dash (---) indicates that it is not,
in which case an NPA calculation adopting the corresponding $SD$, $SDG$, or $SDGI$ pairs
does not reproduce the full collectivity of the ground-state band in the fermionic SU(3) model.
For $n=4$ and $n=8$ nucleons in the $sdg$ shell
no exact mapping can be realized to $sdgi$-IBM
and bosons with even higher angular momentum are needed.
It should be noted, however, that this generally occurs for low nucleon number
(e.g., for $n=12$ nucleons in the $sdg$ shell the problem does not occur),
for which NPA calculations with high angular momentum pairs are still feasible.

\begin{table}
	\caption{
		\label{table1}
		Leading SU(3) representations for 6 bosons in \mbox{$sd$-,} $sdg$-, and $sdgi$-IBM
		occurring in the ${\rm U}( \Lambda)$ and ${\rm U}_{st}(6)$ representations $[\bar h]$
		containing the favored supermultiplet $(0,0,0)$.}
	\begin{tabular}{c|c|l}
		\hline\hline
		(bosons)$^{n_b}$&$[\bar h]$&$(\lambda,\mu)$\\
		\hline 
		$(sd)^6$&$[6]$&$(12,0),(8,2),(4,4),(6,0),(0,6),\dots$\\
		&$[4,2]$&$(8,2),(6,3),(7,1),(4,4)^2,(5,2),\dots$\\
		&$[2^3]$&$(6,0),(0,6),(3,3),(2,2)^2,(0, 0)$\\
		&$[1^6]$&$(0, 0)$\\
		\hline 
		$(sdg)^6$&$[6]$&$(24,0),(20,2),(18,3),(16,4)^2,(18,0),\dots$\\
		&$[4,2]$&$(20,2),(18,3),(19,1),(16,4)^3,(17,2),\dots$\\
		&$[2^3]$&$(18,0),(15,3),(12,6),(13,4),(14,2)^3,\dots$\\
		&$[1^6]$&$(12,0),(8,5),(9,3),(3,9),(7,4),\dots$\\
		\hline 
		$(sdgi)^6$&$[6]$&$(36,0),(32,2),(30,3),(28,4)^2,(30,0),\dots$\\
		&$[4,2]$&$(32,2),(30,3),(31,1),(28,4)^3, (29,2)^2,\dots$\\
		&$[2^3]$&$(30,0),(27,3),(24,6),(25,4),(26,2)^3,\dots$\\
		&$[1^6]$&$(24,0),(20,5),(21,3),(18,6),(19,4),\dots$\\ 
		\hline\hline
	\end{tabular}
\end{table}

\begin{table}
	\caption{
		\label{table2}
		Leading fermionic SU(3) representations $(\lambda,\mu)$
		for $n$ nucleons in the $pf$ and $sdg$ shells
		and the U(6), U(15), and U(28) representations of the $n_{\rm b}=n/2$ boson system
		that contain this $(\lambda,\mu)$
		in $sd$-, $sdg$-, and $sdgi$-IBM.}
	\begin{tabular}{rc|c|c|c}
		\hline\hline
		(shell)$^n$ &$(\lambda,\mu)$&$sd$-IBM&$sdg$-IBM&$sdgi$-IBM\\
		\hline 
		$(pf)^{4}$&$(12,0)$&---&---& $[2]$\\
		$(pf)^{8}$&$(16,4)$&---&---& $ [4],[2^2] $\\
		$(pf)^{12}$&$(24,0)$&---&$[6]$& $[6],[4,2],[2^3],[1^6]$\\
		$(sdg)^{4}$&$(16,0)$&---&---&---\\	
		$(sdg)^{8}$&$(24,4)$&---&---&---\\	
		$(sdg)^{12}$&$(36,0)$&---&---&$[6]$\\	
		\hline\hline
	\end{tabular}
\end{table}

While the best NPA solutions so far have been found
by a numerically intensive optimization,
it turns out they can also be obtained from a deformed ``intrinsic'' state.
Again consider the 6p-6n system in the $pf$ shell.
An unconstrained Hartree-Fock (HF) calculation
in this single-particle shell-model space~\cite{JohnsonSHERPA}
with a quadrupole-quadrupole interaction
provides us with a HF state
with an axially symmetric quadrupole deformed shape,
a consequence of the spontaneous symmetry breaking~\cite{Nambu} of rotational symmetry.
One can project out a $K=0$ band with good angular momentum from this HF state~\cite{JohnsonLAMP},
which exactly corresponds to the SU(3) representation $(24,0)$~\cite{Elliott58}.
We use $a$ and $\bar{a}$ to denote the HF single-particle orbit
and its time-reversal partner, respectively,
and we write the creation operator of a nucleon as $c_{a}^{\dagger}$. 
A Slater determinant for an even number $2N$ of protons or neutrons
can be written as a pair condensate:
\begin{eqnarray}\label{HF}
\prod_{a=1}^{N} c_{a}^{\dagger} c_{\bar{a}}^{\dagger} |0\rangle = \mathcal{N} \left(  \sum_{  a} v_{ a} ~ c^{\dagger}_{a} c_{\bar{a}}^{\dagger} \right)^{N} |0\rangle.
\end{eqnarray}
The pair in the deformed HF state is a superposition
of collective pairs of good angular momentum in the shell model~\cite{arXiv}:
\begin{eqnarray}\label{sp}
 \sum_{  a} v_{ a} ~ c^{\dagger}_{a} c_{\bar{a}}^{\dagger} 
= \sum_{ JM} {{A}^{(J)}_M}^{\dagger}.
\end{eqnarray}
For the appropriate $v_{a}$ one obtains $SDG$ pairs,
which are the same as the $SDG$ pairs obtained by the CG-NPA calculations.
Similarly, the $SDGI$ pairs responsible for (36,0) for 6p-6n in the $sdg$ shell
can be also projected out from a deformed HF pair.
The CG approach provides numerically optimal solutions in the NPA
but is computationally heavy due to hundreds, even thousands of iterations.
The HF approach derives pairs using an unconstrained HF calculation
and the decomposition of pairs according to Eq.~(\ref{sp}) has a very low computational cost.

\begin{figure}
	\includegraphics[width = 0.26 \textwidth]{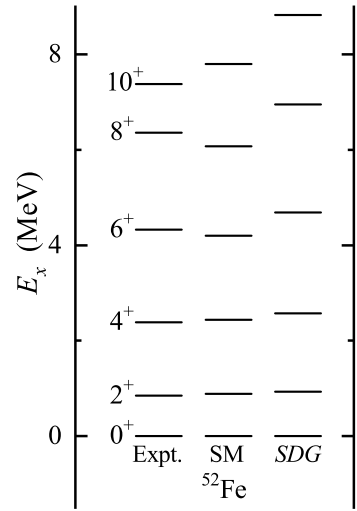}
	\caption{\label{fig3}
The ground rotational band of $^{52}$Fe.
The experimental energies are taken from Ref.~\cite{expt1}
and the shell-model results are obtained with the GXPF1 interaction.}
\end{figure}

\begin{table}\center	
	\begin{tabular}{|c|c|c|c|}  \hline\hline
		$I^{\pi}$  & Expt. & SM &  $SDG$    \\  \hline
		$2^+$ & 14.2(19) & 19.2 & 17.0 \\
		$4^+$  & 26(6)    & 25.0 & 21.6 \\
		$6^+$  & 10(3)    & 17.4 & 20.0 \\
		$8^+$  & 9(4)     & 11.5  & 15.5 \\
		$10^+$ &          & 12.7  & 10.5 \\
		\hline\hline
	\end{tabular}
	\caption{\label{table3}
$B(E2; I \rightarrow I-2)$ values (in W.u.) for the ground rotational band of $^{52}$Fe.
The experimental values are taken from Ref.~\cite{expt1}}
and the shell-model results are obtained with the GXPF1 interaction.
\end{table}

Finally, we show that the NPA with CG-pairs
provides a good description of low-lying states of rotational nuclei
also if a realistic shell-model interaction is taken. 
We exemplify this with the nucleus $^{52}$Fe,
considered as a 6p-6n system in the $pf$ shell with the GXPF1 effective interaction~\cite{gxpf1}.
Figure~\ref{fig3} and Table~\ref{table3} compare, for the ground rotational band of $^{52}$Fe,
the experimental data \cite{expt1}, the full configuration shell model (SM), and the $SDG_{\rm CG}$-pair approximation. 
Both the level energies and the $B(E2)$ values obtained with $SDG_{\rm CG}$
are in good agreement with the data and with the shell model.

In summary, we construct in the NPA
a collective subspace of the full shell-model space
such that the former exactly reproduces, without any renormalization,
the properties of the low-energy states of the latter.
This construction is valid for an SU(3) quadrupole-quadrupole Hamiltonian
and is achieved by determining the structure of the pairs
with the conjugate-gradient minimization technique
or on the basis of a deformed HF calculation.
Exact correspondence is achieved
only if a sufficient number of different pairs  is considered.
For example, a 6p-6n system in the $pf$ ($sdg$) shell
is reproduced exactly with $SDG$ ($SDGI$) pairs; 
with just $SD$ pairs, an important renormalization of all operators is required.
We have analytic understanding of this result:
The collective subspace of the NPA
exactly captures the collectivity of the full space
if and only if the mapping to a model
constructed with bosons corresponding to the pairs
gives rise to a leading bosonic SU(3) representation
that is also leading in fermionic SU(3).

For many years a central problem in nuclear structure
has been the construction of a collective subspace 
that decouples from the full shell-model space.
With this work the conditions necessary for this decoupling to be exact
are now understood for an SU(3) Hamiltonian.
This understanding will pave the way
for the construction of viable collective subspaces
for more realistic shell-model interactions.
It will also clarify the derivation of boson Hamiltonians
appropriate for quadrupole deformed nuclei.
Similar techniques conceivably might be applied elsewhere,
such as to octupole-deformed nuclei
with a Sp($\Omega$) or SO($\Omega$) symmetry~\cite{Isacker2016}.

\begin{acknowledgments}
This material is based upon work supported by
the National Key R\&D Program of China under Grant No. 2018YFA0404403,
the National Natural Science Foundation of China under Grants No. 12075169, 11605122, and 11535004,
the U.S. Department of Energy, Office of Science, Office of Nuclear Physics, under Award No. DE-FG02-03ER41272,
the CUSTIPEN (China-U.S. Theory Institute for Physics with Exotic Nuclei) funded by the U.S. Department of Energy, Office of Science grant number DE-SC0009971.
\end{acknowledgments}


\begin{thebibliography}{100}

\bibitem{Mayer} M. G. Mayer, Phys. Rev. {\bf 75}, 1969 (1949).

\bibitem{Jensen} O. Haxel, J. H. D. Jensen, and H. E. Suess, Phys. Rev. {\bf 75}, 1766 (1949).

\bibitem{BM1} A. Bohr and B. R. Mottelson, Mat. Fys. Medd. K. Dan. Vidensk. Selsk {\bf 27}, 16 (1953).

\bibitem{BM2} A. Bohr and B. R. Mottelson, {\it Nuclear Structure} (World Scientific,1998).

\bibitem{Nilsson} S. G. Nilsson, Mat. Fys. Medd. K. Dan. Vidensk. Selsk {\bf 29}, 16 (1955). 

\bibitem{Rainwater} J. Rainwater, Phys. Rev. {\bf 79}, 432 (1950).

\bibitem{Elliott58} J. P. Elliott, Proc. R. Soc. A {\bf 245}, 128 (1958); {\bf 245}, 562 (1958). 

\bibitem{symmetryadapted} T. Dytrych, K. D. Launey, J. P. Draayer, P. Maris, J. P. Vary, E. Saule, U. Catalyurek, M. Sosonkina, D. Langr, and M. A. Caprio, Phys. Rev. Lett. {\bf 111}, 252501 (2013);
T. Dytrych, K. D. Launey, J. P. Draayer, D. J. Rowe, J. L. Wood, G. Rosensteel, C. Bahri, D. Langr, and R. B. Baker, Phys. Rev. Lett. {\bf 124}, 042501 (2020).

\bibitem{Caurier05}
E. Caurier, G. Mart\'\i nez-Pinedo, F. Nowacki, A. Poves, and A. P. Zuker, Rev. Mod. Phys. {\bf77}, 427 (2005).

\bibitem{Heyde11}
K. Heyde and J. L. Wood, Rev. Mod. Phys. {\bf83}, 1467 (2011).

\bibitem{Otsuka19} T. Otsuka, Y. Tsunoda, T. Abe, N. Shimizu, and P. Van Duppen, Phys. Rev. Lett. {\bf123}, 222502 (2019).

\bibitem{IBM1} A. Arima and F. Iachello,  Phys. Rev. Lett. {\bf 35}, 1069 (1975); Ann. Phys. {\bf 111}, 201 (1978). 

\bibitem{IBM2} F. Iachello and A. Arima, {\it The Interacting  Boson  Model} (Cambridge University Press, Cambridge, 1987).

\bibitem{OAI} T. Otsuka, A. Arima, F. Iachello, and I. Talmi, Phys. Lett. B {\bf 76}, 139 (1978); 
T. Otsuka, A. Arima, and F. Iachello, Nucl. Phys. A {\bf 309}, 1 (1978).

\bibitem{GJ95} J. N. Ginocchio and C. W. Johnson, Phys. Rev. C \textbf{51}, 1861 (1995).

\bibitem{Nomura1} K. Nomura, N. Shimizu, and T. Otsuka, Phys. Rev. Lett. {\bf 101}, 142501 (2008).

\bibitem{Nomura2} K. Nomura, T. Otsuka, N. Shimizu, and L. Guo, Phys. Rev. C {\bf 83}, 041302(R) (2011).

\bibitem{NPA1} J. Q. Chen, Nucl. Phys. A {\bf 626}, 686 (1997).

\bibitem{NPA2} Y. M. Zhao, N. Yoshinaga, S. Yamaji, J. Q. Chen, and A. Arima, Phys. Rev. C {\bf 62}, 014304 (2000).

\bibitem{NPAr} Y. M. Zhao and A. Arima, Phys. Rep. {\bf 545}, 1 (2014).

\bibitem{gs1} I. Talmi, Nucl. Phys. A {\bf 172}, 1 (1971).

\bibitem{bpa1} Y. K. Gambhir, A. Rimini, and T. Weber, Phys. Rev. {\bf 188}, 1573 (1969);
Y. K. Gambir, S. Haq, and J. K. Suri, Ann. Phys.(N.Y.) {\bf 133}, 154 (1981).

\bibitem{bpa2} K. Allaart, E. Boeker, G. Bonsignori, M. Savoia, and Y. K. Gambhir, Phys. Rep. {\bf 169}, 209 (1988).

\bibitem{Lei} Y. Lei, Z. Y. Xu, Y. M. Zhao, and A. Arima, Phys. Rev. C {\bf 80}, 064316 (2009); {\bf 82}, 034303 (2010).

\bibitem{gs2} M. A. Caprio, F. Q. Luo, K. Cai, V. Hellemans, and C. Constantinou, Phys. Rev. C {\bf 85}, 034324 (2012).

\bibitem{gs4} Y. Y. Cheng, Y. M. Zhao, and A. Arima, Phys. Rev. C {\bf 94}, 024307 (2016); 
Y. Y. Cheng, C. Qi, Y. M. Zhao, and A. Arima, Phys. Rev. C {\bf 94}, 024321 (2016).

\bibitem{gs3} C. Qi, L. Y. Jia, and G. J. Fu, Phys. Rev. C {\bf 94}, 014312 (2016).

\bibitem{FDSM1} J. N. Ginocchio, Phys. Lett. B {\bf 79}, 173 (1978); {\bf 85}, 9 (1979); Ann. Phys. {\bf 126}, 234 (1980).

\bibitem{FDSM2} C. L. Wu, D. H. Feng, X. G. Chen, J. Q. Chen, and M. W. Guidry, Phys. Rev. C {\bf 36}, 1157 (1987); C. L. Wu, D. H. Feng, and M. Guidry, Adv. Nucl. Phys. {\bf 21}, 227 (1994).

\bibitem{Halse89} P. Halse, Phys. Rev. C {\bf 39}, 1104 (1989).

\bibitem{Zhao2000} Y. M. Zhao, N. Yoshinaga, S. Yamaji, and A. Arima, Phys. Rev. C {\bf 62}, 014316 (2000).

\bibitem{Xu2009} Z. Y. Xu, Y. Lei, Y. M. Zhao, S. W. Xu, Y. X. Xie, and A. Arima, Phys. Rev. C 79, 054315 (2009).

\bibitem{Lei12un} Y. Lei, S. Pittel, G. J. Fu, and Y. M. Zhao, arXiv: 1207.2297v1;
S. Pittel, Y. Lei, Y.M. Zhao, and G. J. Fu, AIP Conference Proceedings {\bf 1488}, 300 (2012);
S. Pittel, Y. Lei, G. J. Fu, and Y. M. Zhao, Journal of Physics: Conference Series {\bf 445}, 012031 (2013).

\bibitem{CG1} M. R. Hestenes and E. Stiefel, J. Res. Natl. Inst. Stan. {\bf 49}, 409 (1952).

\bibitem{CG2} R. Fletcher and C. M. Reeves, Comput. J. {\bf 7}, 149 (1964).

\bibitem{Wigner} E. P. Wigner, Phys. Rev. {\bf 51}, 106 (1937). 

\bibitem{Elliott81} J. P. Elliott and J. A. Evans, Phys. Lett. B {\bf 101}, 216 (1981).

\bibitem{Isacker2016} P. Van Isacker and S. Pittel, Phys. Scr. {\bf 91}, 023009 (2016).  

\bibitem{Elliott1999} J. P. Elliott and J. A. Evans, J. Phys. G {\bf 25}, 2071 (1999).

\bibitem{JohnsonSHERPA} I. Stetcu and C. W. Johnson, Phys. Rev. C {\bf 66}, 034301 (2002).

\bibitem{Nambu} Y. Nambu, Phys. Rev. Lett. {\bf 4}, 380 (1960).

\bibitem{JohnsonLAMP} C. W. Johnson and K. D. O'Mara, Phys. Rev. C {\bf 96}, 064304 (2017);
C. W. Johnson and C. F. Jiao, Phys. G {\bf 46}, 015101 (2019).

\bibitem{arXiv} G. J. Fu and C. W. Johnson, Phys. Lett. B {\bf 809}, 135705 (2020).

\bibitem{expt1} Y. Dong, H. Junde, Nucl. Data Sheets {\bf 128}, 185 (2015).

\bibitem{gxpf1} M. Honma, T. Otsuka, B. A. Brown, and T. Mizusaki, Phys. Rev. C {\bf 69}, 034335 (2004).

\end{thebibliography}
\end{document}